\documentclass[useAMS,usenatbib]{mn2e}
\usepackage{graphicx}

\def\ucr{${\rm cts\,s^{-1}}$ }
\def\ucre{${\rm cts\,s^{-1}}$}

\def\ergse{${\rm erg\,s^{-1}\,cm^{-2}}$}
\def\ulum{${\rm erg\,s^{-1}}$ }

\def\nh{$N_{\rm H}$ }
\def\nhe{$N_{\rm H}$}

\def\nhexe{$N_{\rm H}^{\rm exc}$}
\def\unh{${\rm cm^{-2}}$ }
\def\unhe{${\rm cm^{-2}}$}

\def\gnh{$N_{\rm H}^{\rm Gal}$ }
\def\gnhe{$N_{\rm H}^{\rm Gal}$}
\def\rxj{RX\,J1028.6-0844 }

\def\gb{GB\,B1508+5714 }
\def\gbe{GB\,B1508+5714}
\def\pmnof{PMN\,J0525-3343 }
\def\pmnofe{PMN\,J0525-3343}
\def\pks{PKS\,B1251-407 }

\def\pmnot{PMN\,J0324-2918 }
\def\pmnote{PMN\,J0324-2918}
\def\pmnft{PMN\,J1451-1512 }
\def\pmnfte{PMN\,J1451-1512}
\def\gbft{GB\,B1428+4217 }
\def\gbfte{GB\,B1428+4217}

\def\asca{{\it ASCA }}

\def\sax{{\it BeppoSAX }}

\def\xmm{{\it XMM-Newton }}
\def\xmme{{\it XMM-Newton}}
\def\chandra{{\it Chandra }}

\def\mnras{\rm MNRAS}
\def\apj{\rm ApJ}
\def\apjl{\rm ApJL}

\def\aj{\rm AJ}
\def\aap{\rm A\&A}

\title[X-ray spectral properties of high-redshift radio-loud quasars]
{X-ray spectral properties of high-redshift radio-loud quasars 
beyond redshift 4---first results \footnote{Based 
on observations obtained with XMM-Newton, an ESA science mission 
with instruments and contributions directly funded 
by ESA Member States and NASA
}}
\author[W. Yuan, et al.]{W. Yuan$^{1}$\thanks{E-mail:
wmy@ynao.ac.cn},
A.C. Fabian$^{2}$,
M.A. Worsley$^{2}$,
and
R.G. McMahon$^{2}$\\
$^{1}$National Astronomical Observatories of China/Yunnan Observatory,
Phoenix Hill, P.O.\ Box 110, Kunming, Yunnan, China\\
$^{2}$University of Cambridge, Institute of Astronomy, 
Madingley Road, Cambridge, CB3 0HA
}

\begin{document}

\date{Accepted Feb 9, 2006, Received ???}

\pagerange{\pageref{firstpage}--\pageref{lastpage}} \pubyear{2005}

\maketitle

\label{firstpage}

\begin{abstract}
We present the results of X-ray spectroscopic observations with \xmm 
for four high-redshift radio-loud quasars  at $z>4$.
Among these, three objects, namely, \gbe,  \pmnote, and \pks
do not show soft X-ray spectral flattening;
the derived upper limits on assumed intrinsic absorption are
$(3.3-17.3)\times 10^{21}$\,\unhe, 
the least of which is 
among the most stringent limits for $z>4$ quasars.
There is a tentative indication for soft X-ray spectral flattening
in \pmnft at z=4.76, though the significance is not high. 
These observations more than double the number of $z>4$ radio-loud
quasars having X-ray spectroscopic data to seven, which compose a 
significant subset of a flux-limited sample of  $z>4$ radio-loud quasars.
Based on this subset we show, in the second part of the paper,
some preliminary results on the overall X-ray spectral properties
of the sample.
Soft X-ray spectral flattening, which is thought to arise from
intrinsic X-ray absorption, was found in about half of the sample
(3/7 or 4/7).
We give  a preliminary distribution of the absorption column density \nhe.
For those with detected X-ray absorption, 
the derived \nh values fall into a very narrow range
(around a few times $10^{22}$\,\unh for `cold' absorption),
suggesting a possible common origin of the absorber.
This \nh distribution is consistent with that 
in the redshift range of 2--4,
though the data are sparse.
Those that do not show X-ray absorption are constrained to have
upper limits on the \nh 
broadly consistent  in general with the lower end of 
the distribution of the detected \nhe.
Compared to lower-redshift samples at $z<2$, there is 
an extension, or a systematic shift, toward higher values
in the intrinsic \nh distribution at $z>4$,
and an increase of the fraction of radio-loud quasars showing 
X-ray absorption toward high redshifts.
These results indicate a  cosmic evolution effect,
which seems to be the strongest at redshifts around 2.
There is a tentative tendency that 
objects showing  X-ray absorption have X-ray fluxes 
systematically  higher than those showing  apparently no absorption.
After the spectral flattening is accounted for,
the rest frame 1--50\,keV continua have photon indices with a mean 
of 1.64 and a standard deviation of 0.11 (or a mean of 1.67 and a 
standard deviation of 0.14 for a Gaussian fit).
Variability appears to be common 
on timescales from a few months to years in the quasar rest-frame,
sometimes in both fluxes and spectral slopes.
\end{abstract}


\section{Introduction}
Quasars with powerful radio emission
are very rare in the early universe.
Beyond  redshift 4, only about a dozen quasars 
with 5\,GHz fluxes brighter than 50\,mJy are known so far
(e.g.\ Hook \& McMahon 1998, Hook et al.\ 2002, Snellen et al.\ 2002).
X-ray spectroscopic observations on a few objects 
revealed some interesting features.
Among these is the flattening of soft X-ray spectra
towards low energies as observed in three objects
\gbfte, \pmnofe, \rxj 
(e.g.\ Boller et al.\ 2000, Yuan et al.\ 2000, 2005, 
Fabian et al.\ 2001ab, Worsley et al.\ 2004ab).
Tentative evidence for the spectral flattening 
was also found in the combined spectra of
several $z>4$, moderately radio-loud quasars with 
\chandra (Bassett et al.\ 2004),
but not in their radio-quiet counterparts 
(e.g.\ Vignali et al.\ 2003, 2005; Grupe et al.\ 2006).
In the lower redshift range $2<z<4$,
this effect was also seen with \xmm observations 
in several highly radio-loud quasars in,  
e.g.\ PKS\,2126-0158 at $z$=3.27
(Ferrero \& Brinkmann 2003) and
RBS\,315 at $z$=2.69 (Piconcelli \& Guainazzi 2005),
and some others as reported recently by Page et al. (2005)
using archival data of \xmme.
In fact, similar results, though subject to relatively large
uncertainties, had already been obtained with other X-ray astronomical
satellites prior to \xmm 
(e.g.\ Wilkes et al.\ 1992; Elvis et al.\ 1994, 1998; Cappi et al.\ 1997;
Brinkmann et al.\ 1997; Fiore et al.\ 1998, 2003; 
Yuan \& Brinkmann 1999; Reeves \& Turner 2000). 
This effect is preferably explained as photoelectric absorption
of soft X-rays (excess absorption), 
though an intrinsic spectral break cannot be ruled out.
When fitted with  `cold' absorption models at
the quasar redshifts, the column densities are around a few times
$10^{22}$\,\unh as measured in recent \xmm observations\footnote{
Earlier observations with \asca and \sax gave somewhat higher column
densities; in this paper we take the \xmm results whenever available,
though the disagreement is not fully understood yet.
}.
The nature of the absorbers (if indeed due to absorption) is not
understood yet. In fact, it is not even known whether this effect is 
ubiquitous in objects of this kind.
Another feature is that a few objects resemble blazars in many ways
(e.g.\ Fabian et al.\ 1997, 1998),
characterized by (apparently) very high X-ray luminosities and
short timescale variability, in which the relativistic beaming effect
plays a role.
However, these results may not be representative of the population
as only a few objects were sampled;
more X-ray spectroscopic data  are needed.

Furthermore, a sizable sample at $z>4$ with good X-ray spectroscopic data 
will provide a basis for the studying of evolution of radio-loud quasars
over a large span of cosmic look-back time. 
This is particularly relevant in the light of recent realization that
at redshifts as high as 4 the cosmic microwave background radiation 
may start to dominate the seed photon field for inverse Compton
processes responsible for the X-ray emission (Schwartz 2002).
Such an example may have been seen  as an X-ray jet/blob associated
with one of these  objects \gb at $z=$4.3 as discovered recently 
with Chandra
(Yuan et al.\ 2003, Siemiginowska et al.\ 2003).

We compiled a flux-limited ($f_{\rm 5Ghz}>50\,mJy$) sample 
of radio-loud quasars at $z>4$, which comprises of about a dozen objects, 
from recent high-$z$ radio quasar surveys 
(e.g.\  Hook \& McMahon 1998, Hook et al.\ 2002, Snellen et al.\ 2002).
In this paper we, firstly, 
report on \xmm observations of several more objects selected from the sample;
secondly, discuss briefly the X-ray spectral properties of the sample  
by combining the new data with those published previously.
It should be noted that these objects (listed in Table\,\ref{tab:sample}) 
represent a significant subset of the whole sample.
The majority of the sample are highly radio-loud quasars 
(radio-loudness\footnote{Defined as the $K$-corrected  ratio of radio (5\,GHz)
to optical ($B$-band) fluxes.} $\gg 10^2$), as compared to those
moderately radio-loud quasars at similar redshifts
studied by Bassett et al.\ (2004) with Chandra, 
in which rather tentative evidence for intrinsic X-ray 
absorption was claimed.
We adopt $H_0$=71\,km\,s$^{-1}$\,Mpc$^{-1}$,
$\Omega_{\Lambda}$=0.73, and $\Omega_{\rm m}$=0.27.
Errors and upper limits are quoted at the 68 per cent level 
unless stated otherwise.

\begin{table*}
\begin{center}
      \caption[]{The sample of $z>$4 radio-loud quasar observed with \xmm}
         \label{tab:sample}
         \begin{tabular}{lcccccl}
            \hline 
            \noalign{\smallskip}
   object  &  RA   & Dec     &  redshift   & $\rm S_{5GHz}$ & magnitude & references \\
                &    \multicolumn{2}{c}{J2000} & &    mJy   & mag & \\                  
            \noalign{\smallskip}
            \hline
            \noalign{\smallskip}
PMN J1451$-$1512 & 14 51 47.05 & $-15$ 12 20.0  & 4.76 &  90 & 19.1($R$) & Hook et al.\ 2002\\
PMN J0324$-$2918 & 03 24 44.28 & $-29$ 18 21.1  & 4.63 & 354 & 18.60($R$) & Hook et al.\ 2002\\
PKS B1251$-$407  & 12 53 59.53 & $-40$ 59 30.7  & 4.46 & 220 & 19.9 ($i$) & Shaver et al.\ 1996\\
GB B1508+5714    & 15 10 02.92 & $+57$ 02 43.4  & 4.30 & 282 & 19.89($i$) & Hook et al.\ 1995\\
\hline
 \multicolumn{7}{c}{Objects with \xmm spectroscopic data published previously}\\
\hline
\gbft            & 14 30 23.74 & $+42$ 04 36.5  & 4.72 & 337 & 20.9      & Worsley et al.\ 2004b\\
\pmnof           & 05 25 06.17 & $-$33 43 05.5   & 4.41 & 210 & 18.7      & Worsley et al.\ 2004a\\
\rxj             & 10 28 38.70 & $-$08 44 38.8   & 4.27 & 159 & 21.0      & Yuan et al.\ 2005\\
            \hline 
         \end{tabular}
\begin{list}{}{}
\item[$^{\mathrm{a}}$] For PMN\,J1451$-$1512 and  PMN\,J0324$-$2918 the positions refer to their
APM {\it R} band optical positions (Hook et al.\ 2002). For PKS B1251$-$407 the radio VLBI position
(Ma et al.\ 1998) is quoted. 
\end{list}
\end{center}
\end{table*}

\section{Results of new X-ray observations}
\subsection{Observations and data reduction}
\label{sect:obs}
The observation logs are summarised in  Table\,\ref{tab:xmmobs}.
Most of the new observations were taken in the \xmm AO3 period,
while \gb was observed previously in 2002.  
\pmnft was observed twice at 5 month apart.
All the EPIC (European Photon Imaging Camera) cameras
 MOS1, MOS2, and PN were active in the observations.
The \xmm Science Analysis System (SAS, v.6.1) and 
the most up-to-date calibration data\footnote{
After we finished the work the new version of SAS (v.6.5)  was released.
We examined any possible effects of the new SAS version
on our results but found no systematic inconsistency;
the changes in the fitted spectral parameters 
are negligible compared to the statistical uncertainties of the measurement.
} (as of July 2005)
were used for data reduction.
We followed the standard data reduction and screening procedures. 
Exposure periods which suffered from
high flaring background caused by soft protons were removed.
The observation for \pks was affected most, and 
much higher cut-off thresholds for background flares than the
recommended values 
(see Loiseau 2004)
\footnote{1\,\ucr and 0.35\,\ucr for
PN and MOS detectors, respectively; see Loiseau 2004.} 
were adopted to ensure enough source counts for spectral analysis.

For all the objects  the X-rays 
were detected at the quasar positions.
Source counts were extracted from a circle of 32\,\arcsec radius
(corresponding to the $\simeq$\,87 per cent encircled energy radius).
Background events were extracted from
source-free regions using a concentric annulus for the MOS detectors,
and one or more circles of  32\,\arcsec radius at
the same CCD read-out column as the source position for the PN detector.
X-ray images, light-curves, and spectra were generated from
the extracted, cleaned events for the source and background.
The data screening criteria and 
the source count rates are given in  Table\,\ref{tab:xmmobs}.
The spectra were created using X-ray events of pattern 0--4 for
PN and 0--12 for MOS, respectively, in the 0.2--8\,keV band, 
and were re-binned to have a minimum of 25--30 counts in each bin.
The EPIC response matrices ({\it rmf} and {\it arf}) were generated
using the source  information on the detectors.

   \begin{table*}
   \begin{center}
      \caption[]{\xmm observation logs, data screening criteria, and source detection}
         \label{tab:xmmobs}
         \begin{tabular}{lccccccc}
            \hline 
            \noalign{\smallskip}
   object (observation) &  obs. id & date of obs. & filter & duration  & bgd cutoff & good exposure & count rate \\
               &          &              &        &  ks       &   \ucre           &   ks          & $10^{-2}$\ucr (0.2--8keV)  \\
               &          &              &        & MOS/PN    & MOS/PN            & MOS/PN        &   MOS/PN  \\
            \noalign{\smallskip}
            \hline
            \noalign{\smallskip}
PMN J1451$-$1512 & 0204190101 & 2004-02-15 & Medium&17.7/14.3 & 0.35/1.2 & 6.8/5.2   & 1.5$\pm$0.2/6.6$\pm$0.4\\
PMN J1451$-$1512 & 0204190401 & 2004-07-22 & Medium&21.1/19.5 & 0.5/1.5  & 19.7/12.0 & 1.7$\pm$0.1/6.1$\pm$0.3\\
PMN J0324$-$2918 & 0204190201 & 2004-02-09 & thin & 13.1/11.5 & 0.35/1.0 & 5.8/4.6   & 1.1$\pm$0.2/4.4$\pm$0.3\\
PKS B1251$-$407  & 0204190301 & 2004-07-12 & thin & 21.8/19.6 & 3.0/15.0 & 17.6/10.5 & 1.3$\pm$0.2/3.6$\pm$0.5\\
GB B1508+5714    & 0111260201 & 2002-05-11 & thin & 13.2/10.9 & 0.3/1.0  & 12.0/9.3  & 4.2$\pm$0.2/15.6$\pm$0.4\\
            \noalign{\smallskip}
            \hline 
         \end{tabular}
\begin{list}{}{}
\item[a] For the MOS detectors the vaules quoted are of MOS1 (MOS2 has almost identical values with MOS1). 
\item[b] {\it bgd cutoff} means the cutoff threshold for flaring background caused by soft protons (for events with energy $>$10\,keV). 
\end{list}
\end{center}
\end{table*}

\subsection{X-ray spectra}
\label{sect:xspec}
We used XSPEC (v.11.3.1) for spectral fitting.
For a given observation, the spectra obtained with the two MOS detectors 
were found to be highly consistent with each other, as expected, and thus 
were combined to form one single MOS spectrum. 
Furthermore, the PN and MOS detectors also yielded statistically 
consistent spectra.
Therefore, for a given observation, joint spectral fitting was performed 
to the PN and the combined MOS spectra. 
The results of the spectral fits are summarised in Table\,\ref{tab:specfit}.
For \pmnft
the spectral fits were performed both to
each observation individually and  to the two observations jointly
(1+2 in Table\,\ref{tab:specfit}).

\begin{figure*}
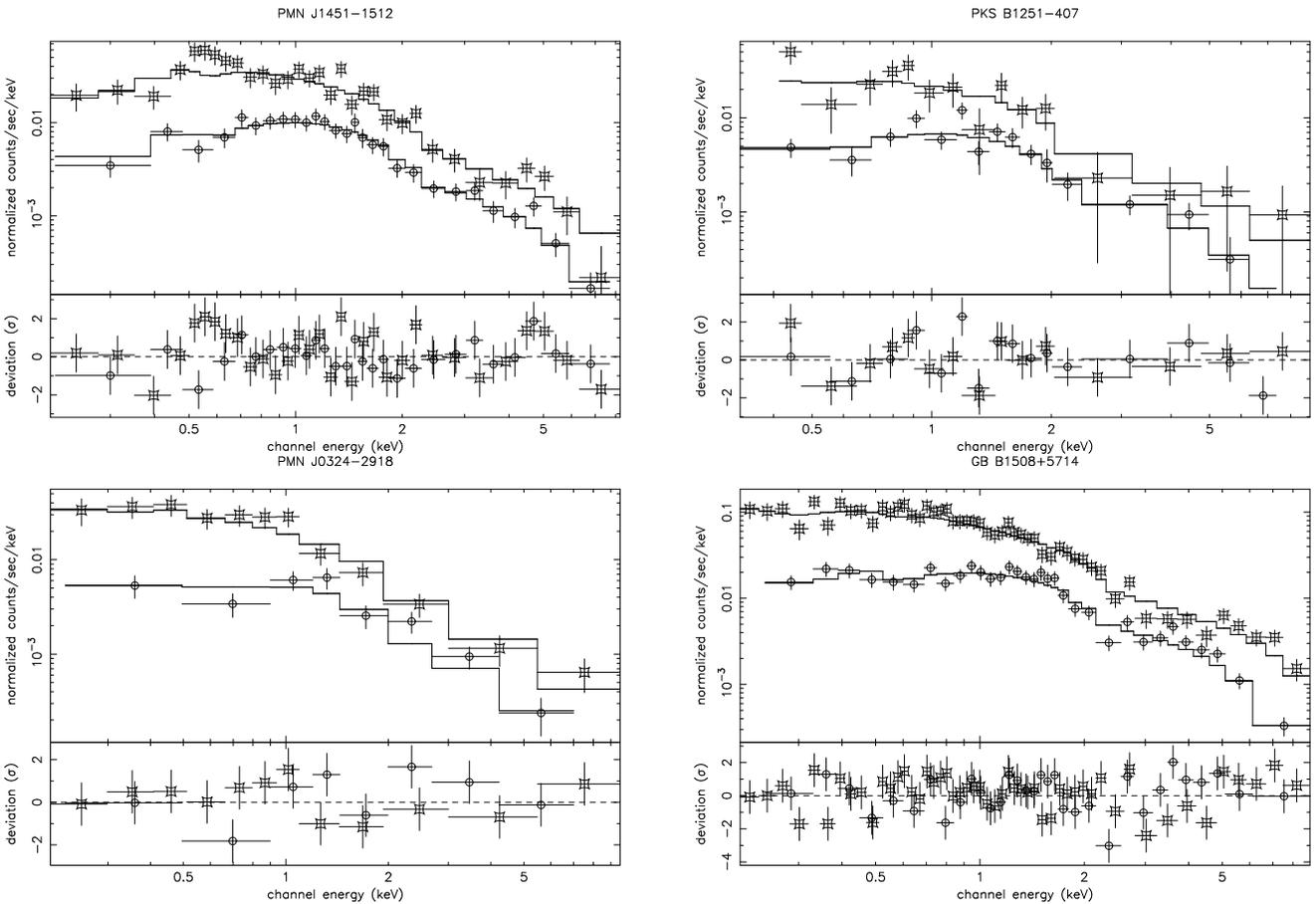

   \centering
   \begin{minipage}[]{0.48\hsize}
   \includegraphics[width=0.7\hsize,angle=-90]{fig1.ps}   
   \includegraphics[width=0.7\hsize,angle=-90]{fig2.ps}
   \end{minipage}
   \hfill
   \begin{minipage}[]{0.48\hsize}
   \includegraphics[width=0.7\hsize,angle=-90]{fig3.ps}
   \includegraphics[width=0.7\hsize,angle=-90]{fig4.ps}
   \end{minipage}
   \caption{\label{fig:fit} The spectra of the four objects measured 
with the PN (squares) and the combined MOS1 and MOS2 (circles)
detectors. The model is the best-fit power-law with Galactic absorption
to the joint PN and MOS spectra, 
and the residuals are the deviations in units of $\sigma$.}
\end{figure*}

\begin{figure*}
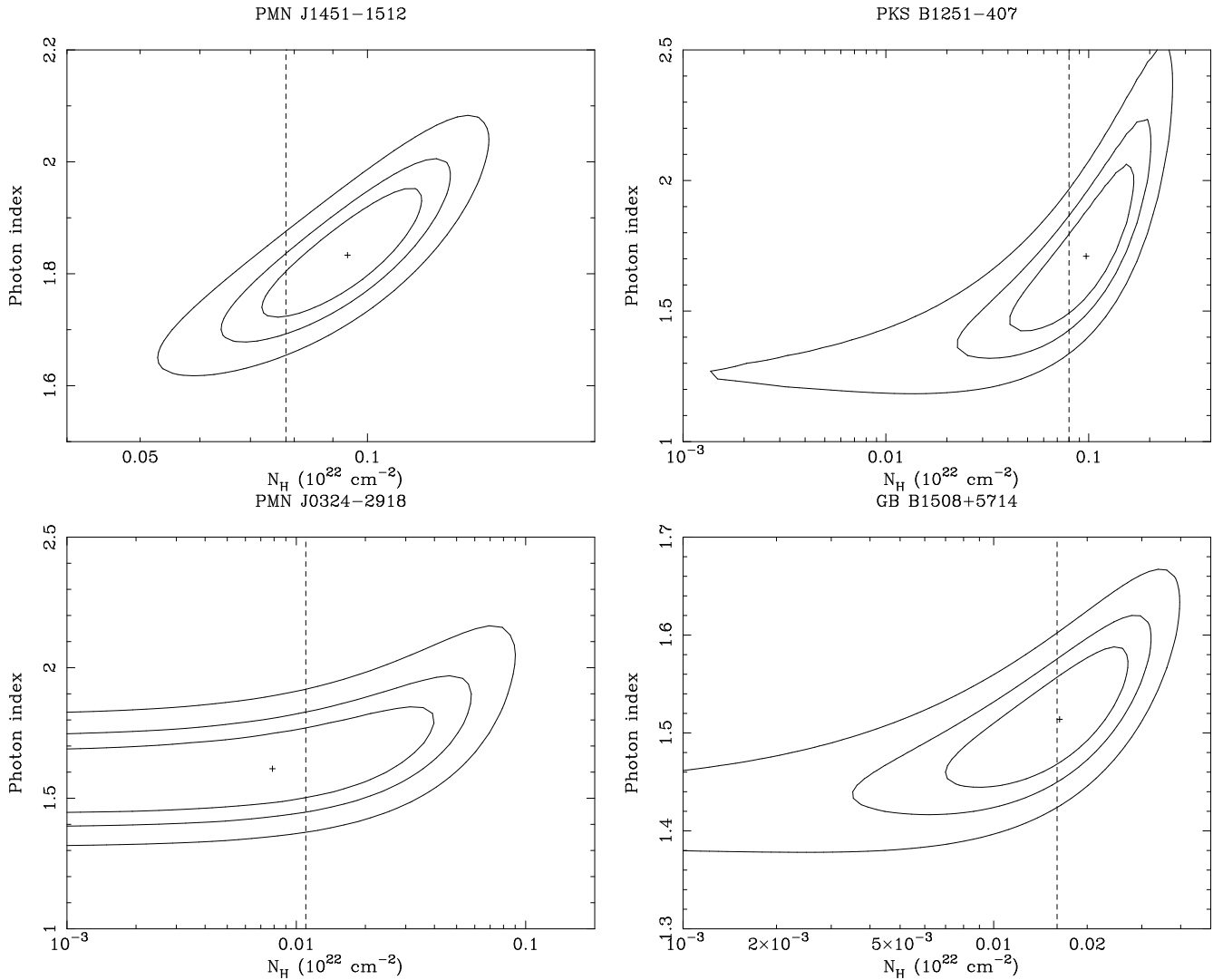

   \centering
   \begin{minipage}[]{0.49\hsize}
   \includegraphics[width=0.82\hsize,angle=-90]{fig5.ps}
   \includegraphics[width=0.82\hsize,angle=-90]{fig6.ps}
   \end{minipage}
   \hfill
   \begin{minipage}[]{0.49\hsize}
   \includegraphics[width=0.82\hsize,angle=-90]{fig7.ps}
   \includegraphics[width=0.82\hsize,angle=-90]{fig8.ps}
   \end{minipage}
   \caption{\label{fig:cont} Confidence contours 
for the fitted spectral photon index and the  
equivalent column density of neutral Hydrogen as 
two interesting parameters. 
The contours correspond to the confidence levels 
of 68, 90, and 99 per cent, respectively.
The cross represents the best-fit values.
The vertical lines indicate the Galactic column densities.
               }
\end{figure*}

All the spectra can be well described with
models of power-law modified by `cold' absorption.
Statistically consistent results were obtained for the fits
with the absorption column density \nh as either 
a free parameter or fixed at the Galactic value 
(Table\,\ref{tab:specfit}).
The X-ray spectra and the best-fit models with \nh fixed at 
the Galactic value are shown in Fig.\,\ref{fig:fit}. 
Plotted in Fig.\,\ref{fig:cont} are 
the confidence contours,  as derived for two interesting  parameters,
for the fitted absorption \nh
in conjunction with the  power law photon indices $\Gamma$.
For \pks and \pmnote, the  low spectral quality 
render relatively large uncertainties for the fitted parameters.
The X-ray fluxes  were calculated by assuming the best-fit
power-law model with Galactic absorption for both the cases
with and without the correction of Galactic absorption.
The fluxes and the quasar rest frame luminosities 
are given in Table\,\ref{tab:flux}.

\begin{table*}
   \begin{center}
      \caption[]{Results of X-ray spectral fits} 
         \label{tab:specfit}
         \begin{tabular}{lccccccccc}
            \hline 
            \noalign{\smallskip}
   object (observation) & \gnh            & $\Gamma$ & $\chi^2$/dof & \nh           &$\Gamma$ & $\chi^2$/dof & \nhexe          & $\Gamma$ & $\chi^2$/dof \\
               & $10^{21}$\,\unh &          &              &$10^{21}$\,\unh&         &              & $10^{21}$\,\unh &          &      \\
            \noalign{\smallskip}
            \hline
            \noalign{\smallskip}
PMN J1451$-$1512 (1)  & 0.78 & 1.78$\pm0.07$ & 20/23 & 0.98$^{+0.24}_{-0.26}$ & 1.87$\pm0.10$ & 20/22 & 11.7$^{+11.9}_{-11.2}$ & 1.87$\pm0.10$& 19/22 \\
PMN J1451$-$1512 (2)  & 0.78 & 1.75$\pm0.05$ & 58/57 & 0.93$\pm0.16$          & 1.82$\pm0.07$ & 57/56 & 10.0$^{+7.2}_{-7.1}$   & 1.83$\pm0.06$& 56/56 \\
PMN J1451$-$1512 (1+2)& 0.78 & 1.76$\pm0.04$ & 78/81 & 0.94$\pm0.13$          & 1.83$\pm0.05$ & 77/80 & 10.4$^{+6.1}_{-5.8}$   & 1.84$\pm0.05$& 75/80 \\
PMN J0324$-$2918      & 0.11 & 1.64$\pm0.09$ & 17/17 & 0.08$^{+0.19}_{-0.08}$ & 1.61$^{+0.15}_{-0.11}$& 17/16& $<$6.5 & 1.64$ ^{+0.11}_{-0.09}$ & 17/16 \\
PKS B1251$-$407       & 0.80 & 1.64$\pm0.11$ & 32/29 & 0.97$\pm$0.38          & 1.71$\pm$0.15 & 32/28 & $<$17.2 & 1.66$\pm$0.13& 32/28 \\
GB B1508+5714         & 0.16 & 1.51$\pm0.03$ & 92/82 & 0.16$\pm$0.06          & 1.51$\pm$0.04 & 92/82 & $<$3.3 & 1.52$\pm$0.04 & 92/81\\
            \noalign{\smallskip}
            \hline 
         \end{tabular}
\begin{list}{}{}
\item[a]The spectral fits were performed jointly to the PN spectrum and the combined MOS1 and MOS2 spectra.
\item[b] \gnhe is the Galactic column density toward 
the direction of a source; \nh is the fitted total local absorption column 
density;  \nhexe refers to the column density of excess absorption;
$\Gamma$ is the power-law photon index.
\end{list}
\end{center}
\end{table*}

\begin{figure}
\includegraphics[angle=-90,width=\hsize]{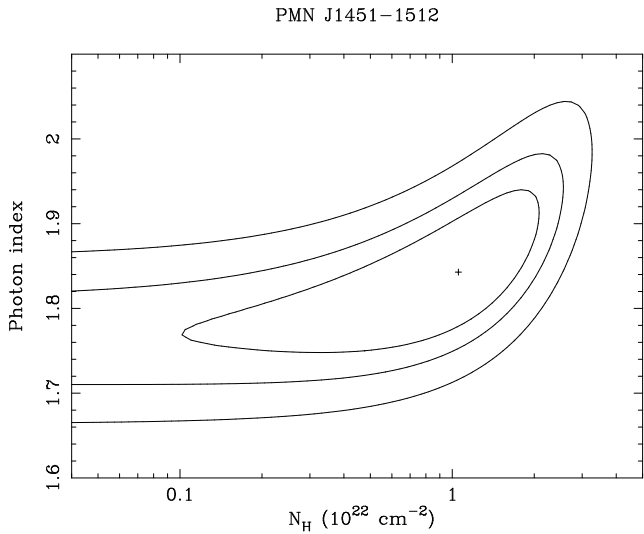}
 \caption{\label{fig:cont_pmn}
Confidence contours for the column density of 
intrinsic absorber versus photon index
for \pmnfte. The fit was performed jointly to the
data from all the detectors from two observations.
The  abundances are assumed to be solar.
The contours correspond to the confidence levels of 68, 90, and 99 per cent,
respectively.
The cross represents the best-fit values.
      }
\end{figure}

The immediate inference from these results is that, 
in general, there is no significant amount of 
excess absorption above the Galactic column density,
though a moderate amount of such absorption cannot be ruled out for 
\pmnfte. 
We thus added a redshifted absorption component
at the quasar redshift in the above model and fixed the local
absorption at the Galactic \gnhe.
This yielded slight improvement in the
fit with $\Delta\chi^2$=3 (for one additional parameter), and 
a best-fit  excess absorption 
\nh= $(10.4^{+6.1}_{-5.8})\times 10^{21}$\,\unhe.
Fig.\,\ref{fig:cont_pmn} shows the confidence contours for the
intrinsic column density and the power-law photon index for \pmnfte. 
It is clear that the statistical significance is not high
and thus the excess absorption is only indicative.
It should be noted that almost identical results were obtained when
fitting the  data set of each individual observation of \pmnfte. 
Assuming the absorber is ionised did not give  a better fit
over the `cold' absorption model.  
For the three remaining objects, no excess \nh is required, 
leading to only upper limits (see Table\,\ref{tab:specfit}). 

   \begin{table}
   \begin{center}
      \caption[]{X-ray fluxes and rest frame luminosities}
         \label{tab:flux}
         \begin{tabular}{lccc}
            \hline 
            \noalign{\smallskip}
   object (observation)    &   flux$^{a)}$   & flux$^{b)}$  & Luminosity \\
               &   \multicolumn{2}{c}{0.2--10\,keV} & 1--50\,keV \\
               &   \multicolumn{2}{c}{$10^{-13}$\,\ergse}  & $10^{46}$\,\ulum  \\
            \noalign{\smallskip}
            \hline
            \noalign{\smallskip}
PMN J1451$-$1512 (1)  & 2.00  & 2.50 & 5.82\\
PMN J1451$-$1512 (2)  & 2.12  & 2.61 & 6.07\\
PMN J0324$-$2918      & 1.32  & 1.39 & 2.98\\
PKS B1251$-$407       & 1.66  & 1.97 & 3.93\\
GB B1508+5714         & 5.29  & 5.55 & 10.2\\
            \noalign{\smallskip}
            \hline 
         \end{tabular}
\begin{list}{}{}
\item[$^{\mathrm{a}}$] Fluxes measured by using the fitted power-law model
in the presence of Galactic absorption.
 \item[$^{\mathrm{b}}$] Same power-law model with a) but with  Galactic absorption corrected.
The values quoted are the mean of the MOS and PN results. 
\end{list}
\end{center}
\end{table}

\subsection{X-ray variability}
\label{sect:vari}
\begin{figure}
\includegraphics[angle=0,width=\hsize]{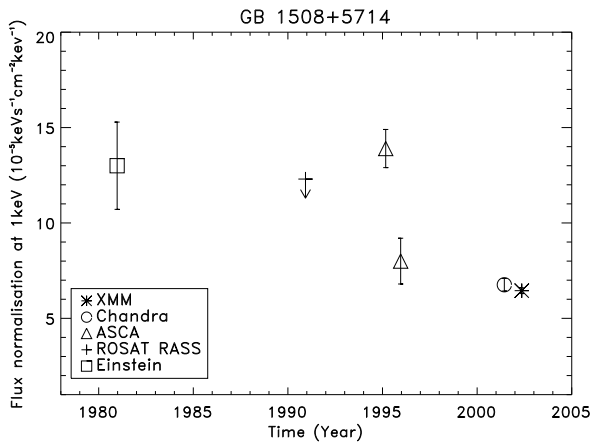}
 \caption{\label{fig:lc_gb}
Long-term light curve for \gb (as the flux normalisation at 1\,keV)
shows strong variability.
The arrow denotes an upper limit placed by the RASS.
The previous data were taken from observations
with {\it Einstein} 
(Mathur \& Elvis 1995), {\it ASCA} (Moran \& Helfand 1997), 
and {\it Chandra} (Yuan et al.\ 2003, Siemiginowska et al.\ 2003). 
}
\end{figure}

The X-ray light curves of the sources
were extracted within the observational intervals 
which were free from background flares.
None of the objects show significant variability on such timescales as
short as about 10\,ksec, i.e.\ $\sim2$\,ksec in the quasar rest frames.
For \pmnft and \gbe, opportunities for exploring
long-term variability are provided by multi-epoch observations and/or 
by previous missions.
The two observations of \pmnfte, separated by 5 months,
show no variability in either flux density or spectral index
over the intrinsic time span of about one month in the quasar rest frame.
Using the results from the literature and this work,
the long-term light curve for 
\gb was constructed and is shown in Fig.\,\ref{fig:lc_gb}
as  the flux normalisation at 1\,keV.
The quasar was not detected in the ROSAT All-sky Survey (RASS), and
an X-ray flux limit was calculated using the X-ray background image
and the exposure map at the quasar position.
A photon index $\Gamma$ 1.5 is assumed when converting the count rates
into fluxes for both the RASS and {\it Einstein} observations.
Long-term X-ray variability is evident 
on time scales of a few years (quasar rest frame).
The large decrease of the X-ray flux observed with {\it ASCA} 
at 9 months (1.7 month in the quasar rest frame) 
apart was already reported by Moran \& Helfand (1997) previously, 
which was accompanied by spectral flattening 
($\Gamma$ from 1.55 to 1.25).
This  decrease is confirmed by the later {\it Chandra} 
and \xmm observations as reported here.
Compared to the first ASCA observation, 
the photon indices found in the \xmm and \chandra observations
are similar to the ASCA value, but the 
flux density dropped by  about 50 per cent.

\section{Discussion: X-ray properties of $z>4$ radio-loud sample}
We discuss in below the X-ray properties of the
objects listed in Table\,\ref{tab:sample} as a class.
They form a significant subset with good X-ray spectroscopic data
of the flux-limited  sample of radio-loud quasars at $z>4$.

\subsection{Intrinsic X-ray absorption}
\begin{figure}
\includegraphics[angle=0,width=\hsize]{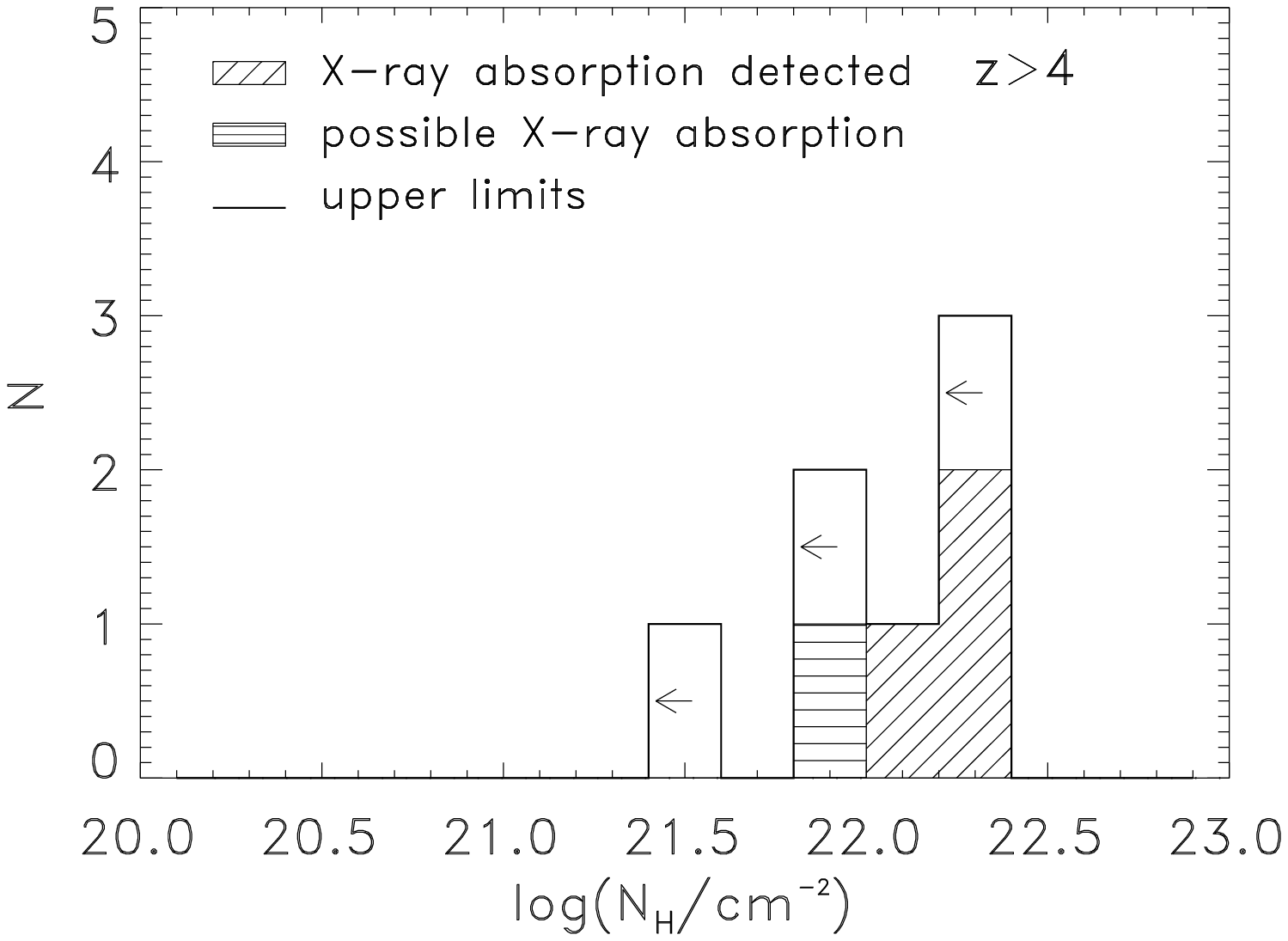}
\includegraphics[angle=0,width=\hsize]{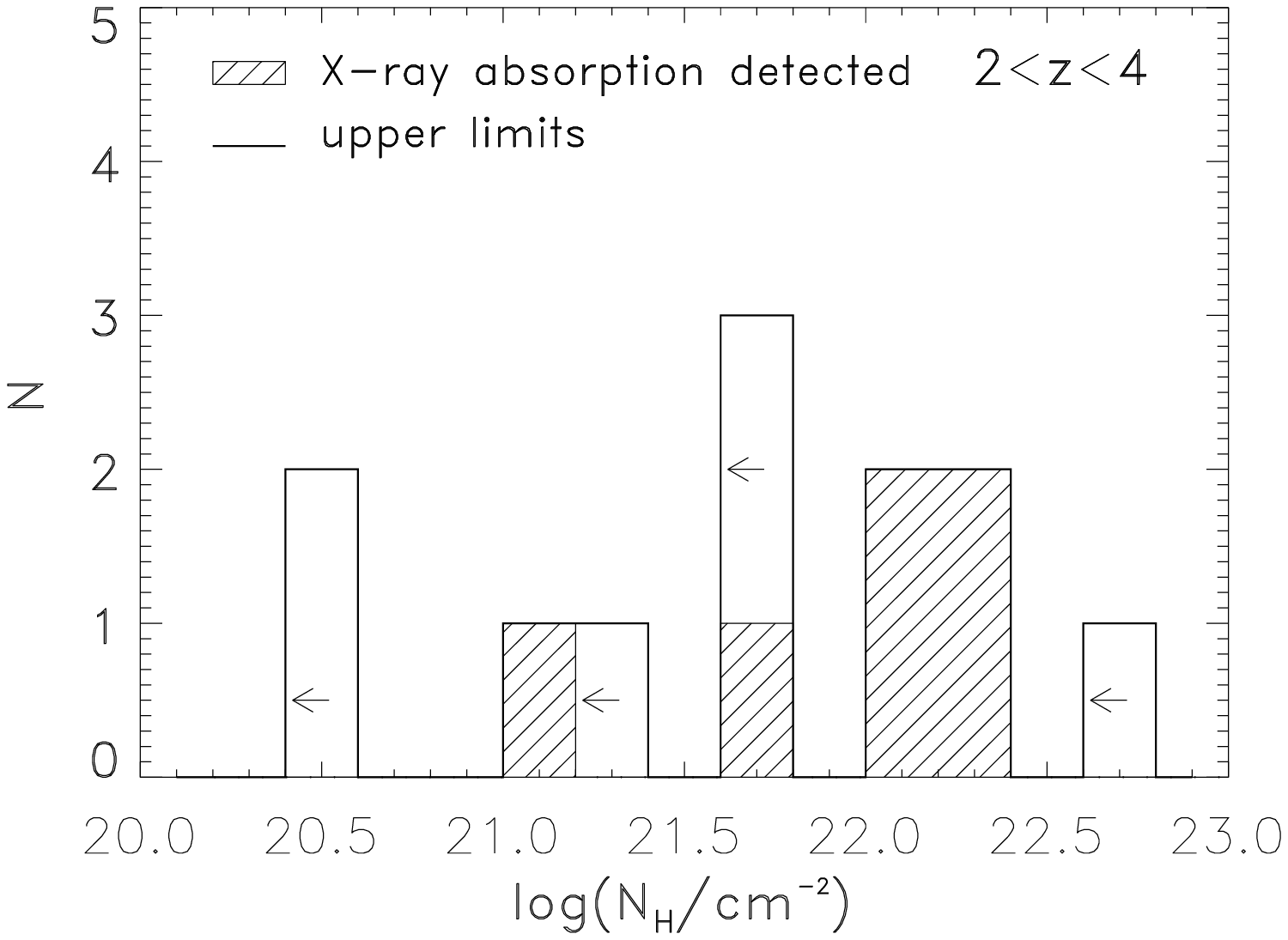}
 \caption{\label{fig:nh_histgm}
Histogram of the equivalent Hydrogen column densities 
for absorber associated
with the quasars (at the quasar redshifts) for the samples of 
 radio-loud quasars with good X-ray spectroscopic data
in the redshift range of $z>4$ (upper panel; from this work)
and of objects in $2<z<4$ (lower panel;  from Page et al.\ 2005).
The detected column densities are represented by hatched regions
while the upper limits (non-detections) by open lines.
Neutral absorption is assumed with the solar metallicity;
the actual column densities are higher if the absorbers are  ionised.
}
\end{figure}

In contrast to the previously detected excess X-ray absorption in 
\gbft (Boller et al.\ 2000, Fabian et al.\ 2001b, Worsely et al.\, 2004b),
\pmnof (Fabian et al.\ 2001a, Worsely et al.\, 2004a), and
\rxj (Yuan et al.\ 2000, 2005), three out of the four quasars reported 
in this paper do not show X-ray absorption down to sensitivity limits 
comparable to, or even lower than, the  \nh values 
found in the three `absorbed objects' above
(a few times $10^{22}$\,\unh
for `cold' absorber).
Tentative evidence for X-ray absorption is suggested in
the remaining object \pmnft
with a \nh value of 1.0$(\pm0.6)\times10^{22}$\,\unhe,
though the significance is not high.
We show in Fig.\,\ref{fig:nh_histgm} (upper panel) the histogram of the
\nh and upper limits of intrinsic X-ray absorption 
derived by assuming neutral absorber.
A few implications can be inferred directly from Fig.\,\ref{fig:nh_histgm},
though the statistics may be affected by the small number of objects 
to some extent.

Firstly, X-ray absorption is detected 
in about half of the sample (3/7 or 4/7).
Secondly, though with sparse data points,
the small dispersion of the derived \nh values for
the detected absorption is remarkable;
for `cold' absorber, they are around a few times $10^{22}$\,\unhe.
The upper-limit for \gb as 3.3$\times10^{21}$\,\unh  
is among the most stringent constraints
on excess \nh obtained so far in high-$z$ quasars 
which do not show  excess absorption.
Yet it is not clear 
whether there exists a distinct sub-class which  shows little or no
X-ray absorption (\nh$\ll 10^{22}$\,\unhe), 
or there is a continuous \nh distribution
with the non-detections falling into the lower \nh end
(see  Fig.\,\ref{fig:nh_histgm}).
The observational difficulty lies in that
at such high redshifts part of the absorption features, 
which are strong at low energy bands, is likely to be
shifted out of the bandpass of the detectors
and renders absorption  with small \nh difficult to detect;
this detection bias cannot be ruled out from the current X-ray data. 
Further X-ray observations for more objects and with better
spectral quality are needed to test these possibilities,
as well as to examine the ionisation parameter of the absorber,
which is poorly constrained in current studies.
If proves to be true, 
the narrow distribution of the absorption \nh
in strong radio-loud quasars at  $z>4$
may suggest a common origin of the absorbers,
which may also share similarities in some other physical properties,
such as metallicity and ionisation parameter.

A comparison of the \nh distribution 
in our $z>4$ sample with those observed at lower redshifts 
could reveal whether and how 
the intrinsic X-ray absorption property evolves with cosmic time.
To make sure that the comparison is free from possible 
cross-calibration problems among different missions 
(see e.g.\ Yuan et al.\ 2005),
we restrict ourselves to results obtained from \xmm
EPIC observations only.
The lower panel in Fig.\,\ref{fig:nh_histgm} shows 
the \nh distribution for a sample of $2<z<4$ highly radio-loud quasars
observed with \xmm collected from the literature 
(Page et al.\ 2005, Ferrero \& Brinkmann 2003, Piconcelli \& Guainazzi 2005).
The results show that,
at least for those with detected absorption,
the \nh distribution in the redshift range 4--5 is
largely consistent with that in $z$=2--4,
though the latter might be somewhat broader than the former.
The fraction  of objects in which X-ray absorption is detected 
is 50 per cent, similar to that in the $z=$4--5 range obtained above
in this work.
These two pieces of evidence imply that the absorber probably underwent 
little or no cosmic evolution in the redshift range from $z$=5 to $\sim2$.

\begin{figure}
\includegraphics[angle=0,width=\hsize]{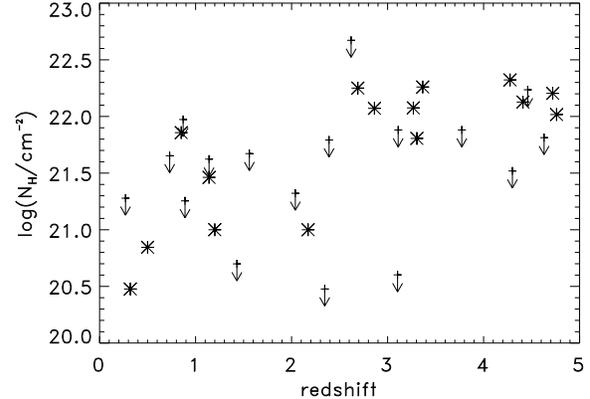}
 \caption{\label{fig:nh-z}
X-ray  absorption column density \nh (assuming neutral gas) 
as a function of redshift.
The data points for $z>4$ objects are taken from this work, 
Yuan et al.\ (2005), and Worsley et al.\ (2004ab); 
those in $2<z<4$ are from  Page et al.\ (2005) and those
in $z<2$  from    Galbiati et al.\ (2005).
For objects without detected X-ray  absorption, the upper limits 
on \nh are  indicated as arrows.
The \nh -- $z$ dependence is significant.
}
\end{figure}

At even lower redshifts (below $z\sim$2), 
we consider the  \nh distribution for a sample
of radio-loud AGN with \xmm data given in Galbiati et al.\ (2005),
the majority of which have $z<2$.
The \nh values given in the Galbiati et al.\ (refer to their Fig.\,5) 
sample are well below\footnote{ 
One object with \nh above  $10^{22}$\,\unh in their figure is 
a type II radio-quiet one and is thus ignored.
It  should be noted that this sample is X-ray selected and is
heterogeneous in terms of the radio power.
}
 $10^{22}$\,\unh 
and are distributed widely in the $10^{19.5-22}$\,\unh range.
Therefore, compared to low redshifts $z<2$, 
an extension or even a systematic shift
toward higher values is indicated in the \nh distribution at redshifts 
above $z\sim2$.
This trend can be seen clearly in Fig.\,\ref{fig:nh-z}, in which the
absorption column densities are plotted versus redshifts
for the objects concerned here.
Exercising a non-parametric correlation test  which takes into
account the upper limits yielded the probability level of 0.005
(Kendall's tau, using the ``Astronomy Survival Analysis'' 
(Isobe, Feigelson \& Nelson 1986)).
A significant \nh --$z$ correlation is thus confirmed
over the whole $z=$0--5 range. 
This is in line with the results claimed previously
in the redshift range of $z<$4 
(Introduction;  see also e.g.\ Page et al.\ 2005).
The increase in \nh with redshift seems to happen at redshifts 
around 2, below and above which no correlation is found from the data, 
respectively.
Furthermore, the fraction of radio-loud quasars showing X-ray absorption
 is much lower at redshifts $z<2$
($<10$ per cent, see e.g.\ Fig.2 in Yuan \& Brinkmann 1999,
see also Brinkmann et al.\ 1997) than the 
high percentage ($\sim 50$ per cent) as found at redshifts above 4 in
this work.
These results, on both the fraction of objects with X-ray absorption  and
the \nh distribution, indicate a strong cosmic evolution effect
in the X-ray absorption property of radio-loud quasars,
which seems to occur at redshifts somewhere around 2.

\begin{figure}
\includegraphics[angle=0,width=\hsize]{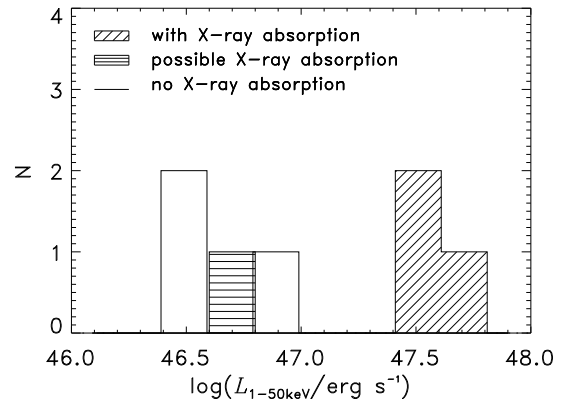}
 \caption{\label{fig:lx_histgm}
Distributions of the 1--50\,keV X-ray luminosities 
(quasar rest frame, assuming isotropic radiation) 
for the sample objects, with their X-ray absorption status indicated.
}
\end{figure}

Fig.\,\ref{fig:lx_histgm} shows the distribution of the apparent 
luminosities (under the assumption of isotropic radiation) in
the 1--50\,keV band for the sample objects, with the different X-ray
absorption properties marked respectively.
Interestingly, there seems to be a systematic difference 
in X-ray luminosities spanning more than one order of magnitude; 
X-ray absorption tends to be 
associated with objects with relatively high luminosities.
It should be noted that this tendency should not be largely
affected by flux variability, since the variability amplitudes
are typically less than a factor of a few.
This could be due, at least partly, 
to the small number of objects used and/or selection effects.
The latter comes into play since,
in general, lower source fluxes result in
poorer X-ray spectra, 
from which X-ray absorption becomes more difficult to detect.
More observational data are needed to improve the statistics and
to eliminate the possible detection biases.
If this trend is true, it may give some insight into the nature of
the X-ray absorbers.

\subsection{Broad band X-ray continuum emission}
\begin{figure}
\includegraphics[angle=0,width=\hsize]{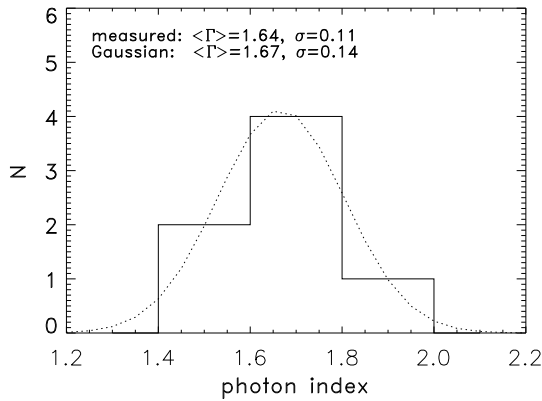}
 \caption{\label{fig:gam_histgm}
Histogram of the power-law photon indices for the sample. 
In the cases where soft X-ray spectral flattening is present,
it is accounted for by intrinsic absorption.
The dotted curve is a fitted Gaussian distribution.
}
\end{figure}

The distribution of the power-law photon indices $\Gamma$ 
(obtained by taking into account excess X-ray absorption, 
if present, see Table\,\ref{tab:specfit})
for the sample is shown  in Fig.\,\ref{fig:gam_histgm}.
The  measurement uncertainties are mostly $\la 0.1$,
smaller than the typical amplitudes of long-term variability 
in spectral slope  ($\Delta\Gamma\sim$0.2--0.3; 
see e.g.\ Moran \& Helfand 1997, Yuan et al.\ 2005).
We find a mean value of $<\Gamma>$=1.64 with 
a standard deviation  $\sigma$=0.11. 
A fit with a Gaussian distribution (dotted curve) gives 
$<\Gamma>$=1.67 and  $\sigma$=0.14.
A comparison with low-$z$ objects (e.g.\ Reeves \& Turner 2000)
indicates little or no
evolution in the continuum spectral slope for radio-loud
quasars from epochs of $z>4$ to the present time.
A tentative  excess emission feature in the rest-frame  5--10\,keV
band was suggested to be  similarly  present in \gbfte, \pmnofe, and \rxj 
(Worsley et al.\ 2004b, Yuan et al.\ 2005);
however, its presence in the four newly observed objects reported here 
could not be tested due to the relatively low  quality of their spectra.

Time variability is  found to be common.
In most objects with more than one observation by either 
\xmm or other instruments, the broad band X-ray fluxes
(1--50\,keV in the quasar rest frame) appear to vary
on timescales of a few months to years in the quasar rest frame,
with amplitudes ranging from 10 percent (e.g.\ \pmnofe)
to a factor of 2 (\gbe). Such variations were often
accompanied by changes in the spectral slopes with
amplitudes of 0.2--0.3 (in \rxj and \gbe), or even higher (in \gbfte).
No variations on timescales shorter than observation intervals,
typically $\sim10$ hours, were observed.
The shortest timescale of variability found was about 7 days
in \gbfte, with an amplitude of $\sim$10 per cent (Worsley et al.\ 2004b).
The apparently extremely high X-ray luminosities 
and their short timescale variability 
suggest a blazar-like nature of these objects,
as discussed in details elsewhere (e.g.\ Fabian et al.\ 1997, 1998;
Moran \& Helfand 1997). 

\section*{Acknowledgements}
WY acknowledges the support by
the National Natural Science Foundation of China (NSF-10533050)
and the BaiRenJiHua programme of 
the Chinese Academy of Sciences.
ACF thanks the Royal Society for their support.
This research has made use of the
NASA/IPAC Extragalactic Database (NED) which is operated by the Jet
Propulsion Laboratory, California Institute of Technology, 
under contract with the National Aeronautics and Space Administration.

\end{document}